\documentclass[12pt]{article}
\usepackage{times}
\usepackage{geometry}
\usepackage[utf8]{inputenc}
\usepackage{enumitem,amssymb}
\usepackage{ragged2e}
\usepackage{graphicx}
\newlist{thematic}{itemize}{8}
\setlist[thematic]{label=$\square$}
\usepackage{pifont}

\begin{document}
\raggedright
\huge
Astro2020 Science White Paper \linebreak

\begin{center}
Discovery of Cold Brown Dwarfs\\ or Free-Floating Giant Planets\\ Close to the Sun
\end{center}
\normalsize

\vskip 0.2in
\noindent \textbf{Thematic Areas:} 
\linebreak
{\it Primary -} $\boxtimes$  Stars and Stellar Evolution  \linebreak
{\it Secondary -} $\boxtimes$ Star and Planet Formation  \linebreak
  
\textbf{Principal Author:}

Name:	Sandy K. Leggett
 \linebreak						
Institution:  Gemini Observatory, Northern Operations
 \linebreak
Email: sleggett@gemini.edu
 \linebreak
Phone:  808 974 2604
 \linebreak
 
\textbf{Co-authors:} 
  \linebreak
  Daniel Apai, University of Arizona\\
  Adam Burgasser, UC San Diego \\
  Michael Cushing, University of Toledo\\
  Trent Dupuy, Gemini Observatory, Northern Operations\\
  Jackie Faherty, American Museum of Natural History\\
  John Gizis, University of Delaware \\
  J. Davy Kirkpatrick, Caltech/IPAC \\
  Mark Marley, NASA Ames \\
  Caroline Morley, University of Texas at Austin\\
  Adam Schneider, Arizona State University\\
  Clara Sousa-Silva, Massachusetts Institute of Technology\\

\vskip 0.5in

\textbf{Description: This White Paper describes the opportunities for discovery of Jupiter-mass objects with 300~K atmospheres. The discovery and characterization of such cold objects is vital for understanding the low-mass terminus of the initial mass function and for optimizing the study of exoplanets by the next generation of large telescopes, space probes and space missions.}

\pagebreak

\begin{center}
{\bf 1. Opportunities} 
\end{center}

\medskip\noindent
The census of free-floating, cold, star-like objects in the neighborhood of our Sun has changed radically every ten years, whenever a new sky survey is launched.
Ten years ago the {\it Wide-field Infrared Survey Explorer} ({\it WISE}, Wri2010) was launched. In the same way that the
far-red  Sloan Digital Sky Survey (SDSS, Yor2000) and the near-infrared Two Micron Sky Survey (2MASS, Skr2006) resulted in the discovery of the L- and T-class dwarfs twenty years ago (Kir1999, Str1999, Leg2000, Bur2006), {\it WISE} images of the sky at $\lambda \approx 5~\mu$m resulted in the discovery of the even cooler and fainter Y-class brown dwarfs  (Cus2011, Kir2012). 
Figure 1 shows the cooling tracks of brown dwarfs and illustrates the evolution of our knowledge of the cold and low-mass population. This white paper describes the opportunities for extending such discoveries to 1 -- 10 Jupiter-mass objects with atmospheres at the freezing point of water.

\begin{figure}[!b]
\begin{center}
\vskip -0.9in
\hskip -0.35in
\includegraphics[angle=0,width=1.05\textwidth]{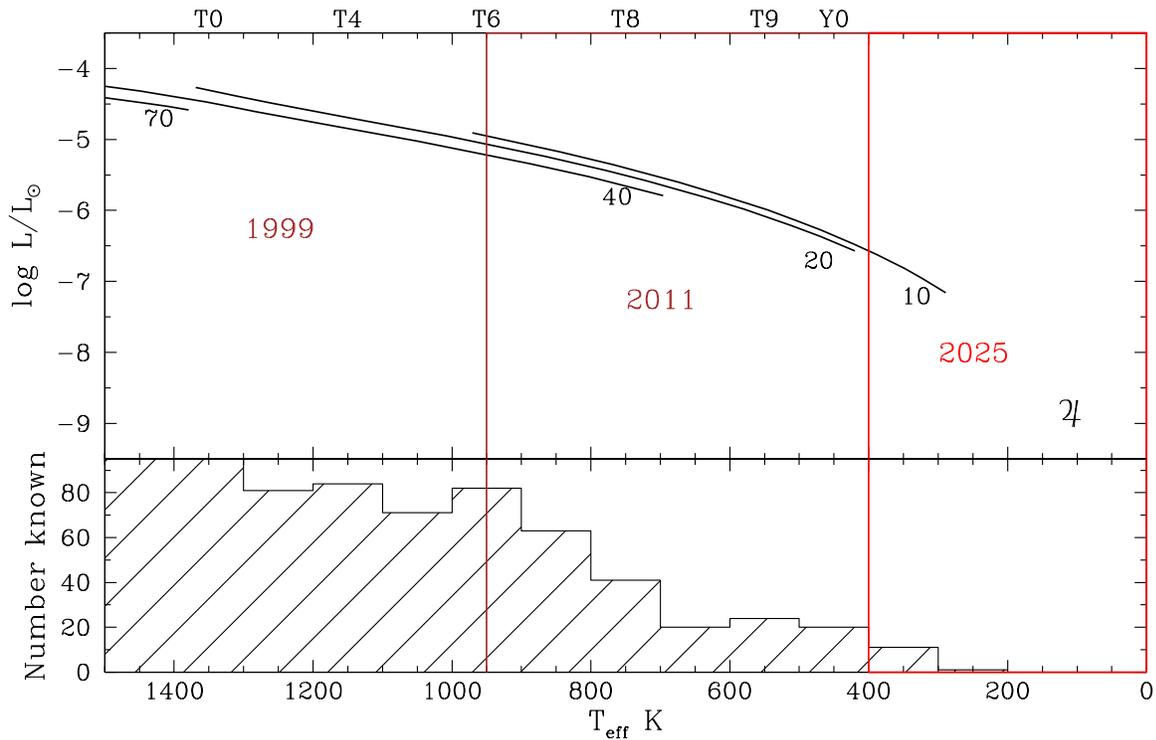}
\vskip -0.45in
\caption{
Upper panel: cooling tracks for brown dwarfs of mass 70, 40, 20 and 10~M$_{\rm Jup}$, for ages 0.1 -- 6~Gyr (Sau2008; black lines). Jupiter is also shown. Spectral types are along the top axis. Temperatures and luminosities probed by 1999 ground-based surveys  (SDSS, 2MASS) and the 2011 space-based survey (WISE) are indicated. Lower:  number of known objects per 100~K.}
\vskip -0.2in
\end{center}
\end{figure}

\medskip\noindent
Theorists have made significant advances in the modelling of cold atmospheres in the last ten years. Molecular linelists have been greatly expanded --- for H$_2$O (Bar2006, Pol2018), NH$_3$ (Yur2011) and CH$_4$ (Har2012; Rey2014; Yur2014). Models now also treat clouds and convection in more physically realistic ways (Mor2012, Mor2014, Tre2015, Hel2017). While models should and can be further improved, for example by creating three dimensional hydrodynamic models, we are in a much better position to analyse cold atmospheres than we were ten years ago.

\medskip\noindent
As the last two decades have taken us to the frontiers of, first, the 1000~K substellar population, and then the 500~K population (Figure 1), in the next decade we have the opportunity to find the 250~K population. One such object is currently known, 2 parsecs from the Sun (Luh2014). They exist and we can find them, if sufficient resources are put to the task.

\medskip
\begin{center}
{\bf 2. Context} 
\end{center}

\begin{figure}[!b]
\begin{center}
\includegraphics[angle=0,width=0.75\textwidth]{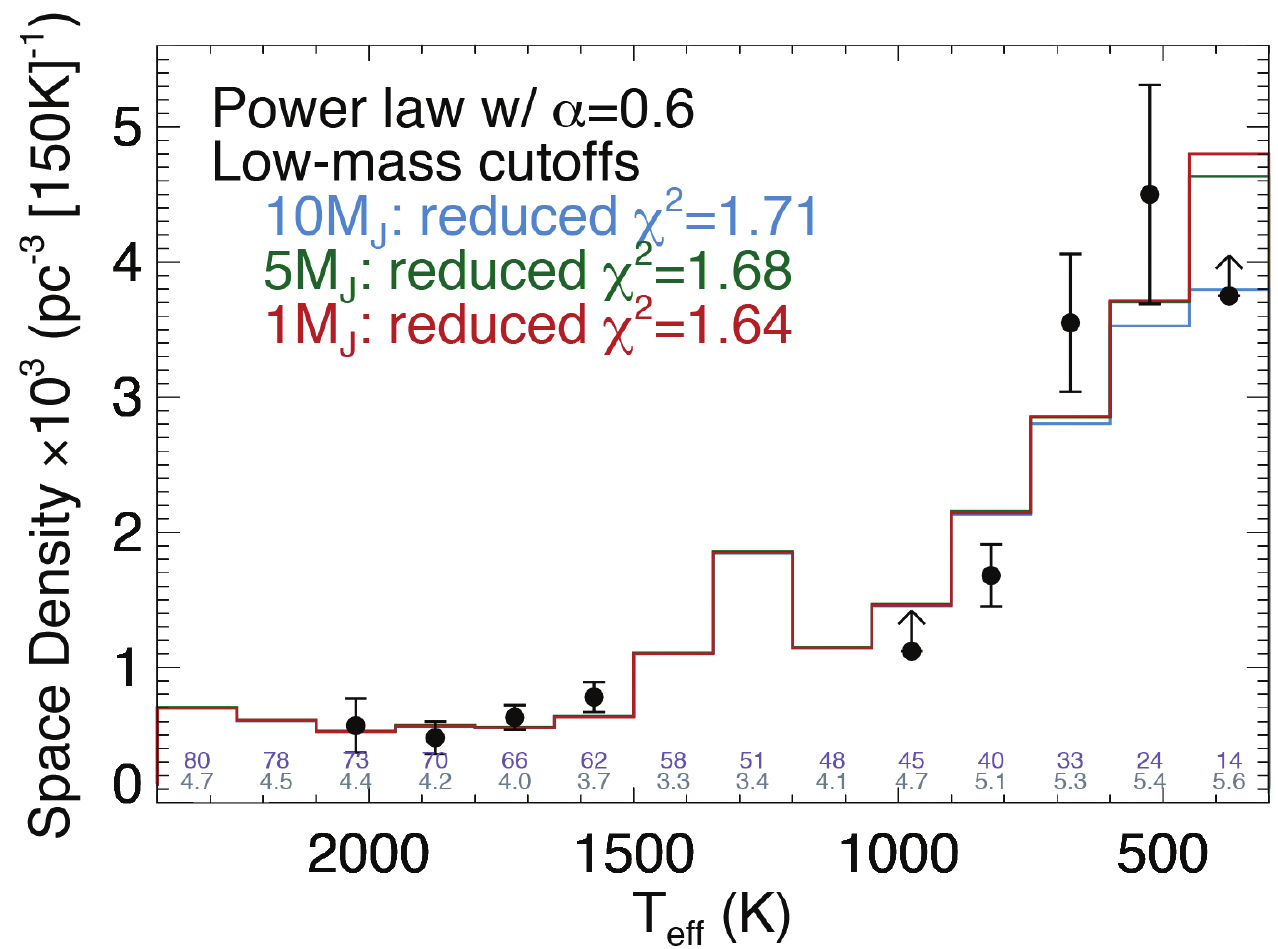}
\caption{Power law fits to the solar neighborhood population of low-mass stars and brown dwarfs (from Kir2019). The best mass function fit gives $\alpha = 0.6$ and a minimum mass of ``star'' formation of 1~M$_{\rm Jup}$.   The mean mass (M$_{\rm Jup}$) and age (Gyr) in each bin is shown along the bottom axis.
}
\end{center}
\end{figure}

\medskip\noindent
Figure 2 shows our current knowledge of the very low-temperature and low-mass Solar-neighborhood population. Current instrumentation limits the detection of cold objects to small distances, and so few are known (Figure 1). However Figure 2 shows that, when expressed as a space density, the number of objects rises as temperature and mass decrease. At the cold limit of the known population, 400~K, the typical object has a mass of 14~M$_{\rm Jup}$ and an age of 5.6~Gyr. 

\medskip\noindent
The best mass function fit to the known Solar-neighborhood population suggests that the sample includes objects as low-mass as 1 -- 5~M$_{\rm Jup}$. This is supported by studies of young clusters and moving groups, where objects with masses $\sim$ 3 -- 6~M$_{\rm Jup}$  have been found (e.g. Liu2013, Gag2015, Fah2016, Bes2017, Esp2017).  Such objects are younger and warmer than the target sample of this paper --- typically  they have temperatures $\sim$1500~K and ages $\lesssim 20$~Myr. We are seeking the older and therefore colder versions of these objects that exist near the Sun. Because all very-low-mass objects ever formed still exist, the number of such objects will constrain models of star formation and of brown dwarf evolution --- at masses and ages hitherto unexplored.

\medskip\noindent
The new nearby $\sim$Jupiter-mass objects will be extrasolar giant planet analogs that will be amenable to detailed investigation. Their atmospheres will provide information on: convection and energy transport, molecular opacities, non-equilibrium processes, cloud formation, grain absorption and scattering, and three-dimensional time-dependent weather. Brown dwarf spectra have long been essential to model processes in both stellar and exoplanetary atmospheres, and have lead to the recognition of new effects such as the role of silicate and sulfide clouds (Mor12). We have the opportunity to explore new territory where objects have low surface gravities that are similar to the giant planets, and have cold atmospheres where H$_2$O and NH$_3$ condense.

\medskip\noindent
Surprises are likely. For example, cool brown dwarfs were found to have a factor of $\sim$2 less flux at $\lambda \approx 4.5~\mu$m than anticipated because 
non-equilibrium chemistry is unexpectedly prevalent in their atmospheres (Sau2007). Similarly, strong absorption by NH$_3$ in the near-infrared was expected for objects cooler than the T dwarfs (e.g. Bur2003), however non-equilibrium chemistry favors N$_2$ over  NH$_3$ and the near-infrared spectra of Y dwarfs are very similar to that of the T dwarfs (e.g. Sch2015). The dramatic change in color across the L- to T-type transition, where volatiles condense and CH$_4$ gas becomes abundant (Bur2002), was not predicted. Likewise, we may see substantial changes in brown dwarf spectra as we reach temperatures where H$_2$O condenses.

\medskip\noindent
The discovery and characterization of the cold brown dwarfs is vital, not only for understanding the makeup of low-mass objects in the local disk, but also for optimizing the study of exoplanets by the next generation of very large telescopes, space probes and space missions.

\medskip
\begin{center}
{\bf 3. Required Observations}
\end{center}

\medskip\noindent
Figure 3 shows the modelled spectral energy distributions of brown dwarfs which are representative of the population we aim to find (Figure 1). Jupiter's emission spectrum if it were at 0.5~pc is also shown, to demonstrate the detectability of a free-floating Jupiter.   Also, the brightest scenarios for the putative Planet 9 (a Uranus-type object at 700~au; Bat2016) indicate that it may be detectable at $\lambda \sim 4~\mu$m with a flux at the Earth of $\sim 1e-18$ ~Wm$^{-2}\mu$m$^{-1}$ (For2016).

\begin{figure}[!t]
\begin{center}
\vskip -0.9in
\includegraphics[angle=-90,width=1.0\textwidth]{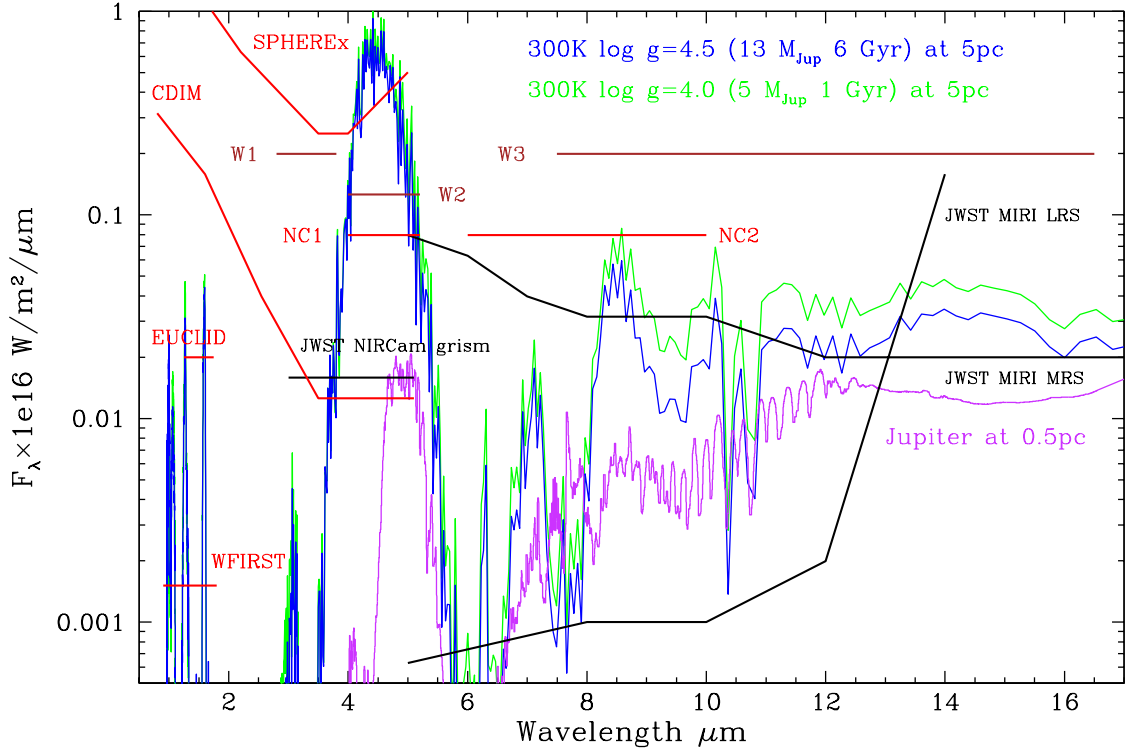}
\vskip -0.25in
\caption{Blue and green lines are synthetic spectra for 300~K brown dwarfs at 5~pc with different mass and age (Tre2015). The purple line is the thermal spectrum of Jupiter at 0.5~pc (Irw2009).  Black lines indicate the limits of the {\it JWST} instruments. Red lines show limits for planned and proposed surveys: CDIM 300~deg$^2$; SPHEREx all-sky; EUCLID 15,000~deg$^2$;
WFIRST 2,200 ~deg$^2$; NEOCam  27,000 ~deg$^2$ (two filters, nominally NC1 and NC2). Brown lines are limits reached by the {\it WISE} all-sky survey in the W1, W2 and W3 filters. Note that the $y$-axis scale is logarithmic.
}
\end{center}
\end{figure}

\medskip\noindent
The dominance of the 4.5~$\mu$m wavelength region for these very cold objects is demonstrated in Figure 3. Figure 3 also illustrates the synergy between the discovery of the icy Jupiters and the availability of the mid-infrared instruments on the {\it James Webb Space Telescope, JWST}. In particular, for objects between the temperature of Jupiter and the early-Y dwarfs, sensitivity at $\lambda = 4$ -- 5~$\mu$m and beyond 8~$\mu$m is critical for determining the physical properties of these objects. There is also synergy with WFIRST which will image exoplanets in reflected light; isolated exoplanet analogs will provide a direct measurement of the thermal flux contribution. 

\medskip\noindent
Figure 3 shows the $5\sigma$ sensitivity limits reached by the {\it WISE} mission (Wri2010) \footnote{http://wise2.ipac.caltech.edu/docs/release/allsky/}. Also shown are the limits for currently planned and proposed near- to mid-infrared surveys: the Cosmic Dawn Intensity Mapper (CDIM, Coo2016 and Unw2019); SPHEREx (Dor2018);  EUCLID (Joa2016); WFIRST (the High Latitude Survey, Spe2015), and  NEOCam (Mai2015, Mas2019).

\medskip\noindent
Currently $\sim$six 300 -- 350~K brown dwarfs are known (Leg2017, Kir2019) out to the {\it WISE} limit of $\sim 12$~pc --- at these cold temperatures changes of 50~K correspond to factors $\sim$2 in luminosity. This number implies a space density of $\sim 1e-3$~pc$^{-3}$ for a 50~K range
which is consistent with the $\approx 4e-3$ 400 -- 550~K objects per pc$^3$ determined by Kir2019 and shown in Figure 2.  Adopting a density of $1e-3$~pc$^{-3}$ per 50~K, Table 1 gives the number of 300~K brown dwarfs that are likely to be found by planned and proposed missions. Between zero and eight 300~K objects will be found by these missions as they are currently specified.

\begin{table}
\begin{minipage}{.5\linewidth}
\footnotesize
\caption{\bf Number of 275--325~K Objects Found}
\smallskip
\centering
\begin{tabular}{lrrr}
\hline \\
Mission/ &  Area & Depth  & Number \\
Probe &  1000 deg$^2$ & pc &  \\
\hline \\
CDIM wide & 0.3 & 40 & 2 \\
EUCLID & 15 &  6  &  0\\
NEOCam  & 27  & 15  & 8 \\
SPHEREx & 41 & 7 & 1 \\
WFIRST & 2.2 & 22 & 2 \\
Proposed Narrow  & 2.5 &  50  & 26  \\
Proposed Wide  & 20 &  25  & 26  \\
Proposed All-Sky  & 41 &  20  & 28  \\
\hline \\
\end{tabular}
\end{minipage}%
\hskip 0.3in
\begin{minipage}{.5\linewidth}
\footnotesize
\centering
\caption{\bf Brightness  at $\lambda =$~4.5~$\mu$m}
\smallskip
\begin{tabular}{lrrrr}
\hline \\
$T_{\rm eff}$ & $M_{[4.5]}$ & $\mu$Jy at &  $\mu$Jy  at&  $\mu$Jy at\\ 
    K     &  Vega  & 20~pc & 25~pc & 50~pc \\
\hline \\
500 & 14.2 & 80  & 50 & 13 \\
450 & 14.5 & 60 & 40 & 10 \\
400 & 14.8 & 40 & 25 & 7 \\
350 & 15.3 & 30  & 20 & 5 \\
300 & 15.9 & 18 & 12 & 3 \\
250 & 16.9 & 6 & 4 & 1 \\
\hline \\
\end{tabular}
    \end{minipage} 
\vskip -0.2in
\end{table}

\medskip\noindent
Tens of objects per 50~K are needed for scientific analyses of these cold objects because they will not be a uniform sample. At a given $T_{\rm eff}$ there will be older more massive brown dwarfs as well as younger lower-mass objects (Figures 1, 3). Studies of the Milky Way show the stars at the position of the Sun to  have ages of 1 -- 7~Gyr and metallicities of $-0.3$ -- $+0.2$ dex (e.g. Mac2017). The small sample of Y dwarfs currently known (Figure 1) reflects this inhomogeneity, having a likely range in metallicity of $-0.5$ -- $+0.4$ dex and in age of 0.5 -- 8~Gyr (Leg2017). Figure 3 shows that although the 300~K objects with different gravities (i.e. mass and age) are similar at 4.5~$\mu$m, the spectra at $\lambda > 8~\mu$m are very different. Calculations show that metallicity differences will also significantly impact the spectra (Tre2015, Leg2017). Hence a handful of spectra cannot be assumed to be representative of the population.

\medskip\noindent
To derive meaningful statistics for 300~K $\sim$Jupiter-mass objects, a minimum of 25 new members of this population is necessary.   Table 1 shows that NEOCam will find the largest number of cold brown dwarfs, however   the volume surveyed would need to be quadrupled, meaning a factor of 2.5 increase in sensitivity. An extended life for NEOCam combined with stacking of images  may give the necessary depth (Kirkpatrick et al. 2019 white paper).

\medskip\noindent
A new mission must probe volumes  $\sim 30000$~pc$^3$, assuming a density of $1e-3$~pc$^{-3}$.  Sources must be within
$\sim$50~pc for followup with {\it JWST} (Figure 3) and so the minimum sky area is $\sim 2500$~deg$^2$. A wider 20000~deg$^2$ area out to $\sim$ 25~pc or an all-sky map out to $\sim$ 20~pc would also satisfy the volume requirement. Table 1 lists these options as proposed Narrow, Wide and All-Sky surveys. Table 2 lists the sensitivities needed to detect cold objects out to 20, 25 and 50~pc.

\medskip\noindent
Only nearby extremely cold objects will be detected at 
wavelengths other than $4.5~\mu$m (Figure 3).  Repeat observations at different epochs are therefore needed to exclude transients and artefacts, and hence the probe or mission will need a $> 1$ year lifetime to map a sufficiently large area more than once. 
Multiple detections will also provide kinematic information and constrain the number of wide planet or brown dwarf stellar companions. Warmer objects will be excluded by their detection at $\lambda < 4.5~\mu$m. Discoveries will be targetted by {\it JWST} for  low-resolution spectra to characterize molecular features; medium-resolution observations (to examine e.g. chemical abundances) would be limited to objects closer than 10~pc (Figure 3).

\medskip
\begin{center}
{\bf 4. Summary}
\end{center}

\medskip\noindent
None of the currently planned or proposed space missions or probes will find the number of cold objects that is required to characterize the low-mass terminus of the Solar-neighborhood mass function. Extension or revision of the planned missions, or design of a new mission, is necessary. The new mission would execute a survey at $\lambda \approx 4.5~\mu$m over the entire sky, half the sky or 6\% of the sky to  sensitivities of  $\sim 12~\mu$Jy, 
$\sim 8~\mu$Jy, or $\sim 2~\mu$Jy, respectively.

\medskip\noindent
The science community has the opportunity  to give the next generation of astronomers  an entirely new population of cold objects to study. Such objects will have masses of a few Jupiters, surface temperatures around the freezing point of water, and lie within tens of parsecs of the Sun. We must enable this exploration.

\begin{table}[!t]
{
\caption{\bf References}
\medskip
\begin{tabular}{ll}
Bar2006 & Barber, R. J., Tennyson, J., Harris, G. J. \& Tolchenov, R. N., 2006, MNRAS, 368, 1087\\
Bat2016 & Batygin, Konstantin \& Brown, Michael E., 2016, AJ, 151, 22 \\
Bes2017 & Best, W. M. J., Liu, M. C., Dupuy, T. J. \& Magnier, E. A., 2017, ApJ, 843, L4\\
Bur2002 & Burgasser, A. J., Marley, M. S., Ackerman, A. S., Saumon, D., 2002, ApJ, 571, L171 \\
Bur2006 & Burgasser, A. J., Geballe, T. R., Leggett, S. K., et al., ApJ, 637, 1067\\
Bur2003 & Burrows, A., Sudarsky, D. \& Lunine, J. I., 2003, ApJ, 596, 587 \\
Coo2016 & Cooray, A., Bock, J., Burgarella, D., Chary, R.,  et al., 2016, arXiv:1602.05178 \\
Cus2011 & Cushing, M. C., Kirkpatrick, J. D., Gelino, C. R., Griffith, R. L., et al., 2011, ApJ, 743, 50\\
Dor2018 & Dor\'{e}, O., Werner, M. W., Ashby, M. L. N., Bleem, L. E., et al., 2018,  arXiv:1805.05489 \\
Esp2017 & Esplin, T. L., Luhman, K. L., Faherty, J. K., Mamajek, E. E., et al., 2017, AJ, 154, 46 \\
Fah2016 & Faherty, J. K., Riedel, A. R., Cruz, K. L., Gagn\'{e}, J., et al., 2016, ApJS, 225, 10 \\
For2016 & Fortney, J. J., Marley, M. S., Laughlin, G., Nettelmann, N., et al., 2016, ApJ, 824, L25\\
Gag2015 & Gagn\'{e}, J., Faherty, J. K., Cruz, Kelle L., Lafreni\'{e}re, D., et al., 2015, ApJS, 219, 33\\
Har2012 & Hargreaves, R. J., Beale, C. A., Michaux, L., Irfan, M. \& Bernath, P. F., 2012, ApJ, 757, 46\\
Hel2017 & Helling, Ch., Tootill, D., Woitke, P. \& Lee, G., 2017, A\&Ap, 603, 123\\
Irw2009 & Irwin, P., 2009, Giant Planets of Our Solar System, Springer Praxis Books\\
Joa2016 & Joachimi, B., 2016, ASP Conference Series, Vol. 507, p.401 \\
Kir1999 & Kirkpatrick, J. D., Reid, I. N., Liebert, J., Cutri, R. M.,  et al., 1999, ApJ, 519, 802\\
Kir2012 & Kirkpatrick, J. D., Gelino, C. R., Cushing, M. C., Mace, G. N.,  et al., 2012, ApJ, 753, 156\\
Kir2019 & Kirkpatrick, J. D., Martin, E. C., Smart, R. L., Cayago, A. J., et al., 2019, ApJS, 240, 19\\
Leg2000 & Leggett, S. K., Geballe, T. R., Fan, X., Schneider, D. P., et al., 2000, ApJ, 536, L35\\
Leg2017 & Leggett, S. K., Tremblin, P., Esplin, T. L., Luhman, K. L. \& Morley, C. V., 2017, ApJ, 842, 118\\
Liu2013 & Liu, M. C., Magnier, E. A., Deacon, N. R., Allers, K. N.,  et al., 2013, ApJ, 777, L20\\
Luh2014 & Luhman, K.L., 2014, ApJ, 786, L18\\
Mai2015 & Mainzer, A., Grav, T., Bauer, J., Conrow, T., et al., 2015, AJ, 149, 172\\
Mac2017 & Mackereth, J. T., Bovy, J., Schiavon, R. P., Zasowski, G.,  et al., 2017, MNRAS, 471, 3057\\
Mas2019 & Masiero, J., Mainzer, A. \& Wright, E. L., 2019, AAS Meeting \#233, id.\#363.10\\
Mor2012 & Morley, C. V., Fortney, J. J., Marley, M. S., Visscher, C.,  et al., 2012, ApJ, 756, 172 \\
Mor2014 & Morley, C. V., Marley, M. S., Fortney, J. J., Lupu, R., et al., 2014, ApJ, 787, 78\\
Pol2018 & Polyansky, O. L., Kyuberis, A. A., Zobov, N. F., Tennyson, J.,  et al.,  2018, MNRAS, 480, 2597\\
Rey2014 & Rey, M., Nikitin, A. V. \& Tyuterev, Vl. G., 2014, ApJ, 789, 2\\
Sau2007 & Saumon, D., Marley, M. S., Leggett, S. K., Geballe, T. R.,  et al., 2007, ApJ, 656, 1136 \\
Sau2008 & Saumon, D. \& Marley, M. S., 2008, ApJ, 689, 1327\\
Sch2015 & Schneider, A. C., Cushing, M C., Kirkpatrick, D., Gelino, C. R., et al., 2015, Ap.J.,  804, 92\\
Skr2006 & Skrutskie, M. F., Cutri, R. M., Stiening, R., Weinberg, M. D.,  et al., 2006, AJ, 131, 1163\\
Spe2015 & 	Spergel, D., Gehrels, N., Baltay, C., Bennett, D., et al., 2015, arXiv:1503.03757 \\ 
Str1999 & Strauss, M. A., Fan, X., Gunn, J. E., Leggett, S. K.,  et al., 1999, ApJ, 522, L61\\
Tre2015 & Tremblin, P., Amundsen, D. S., Mourier, P., Baraffe, I.,  et al., 2015, ApJ, 804, L17\\
Unw2019 & Unwin, S. C., AAS Meeting \#233, id.\#158.04\\
Wri2010 & Wright, E. L., Eisenhardt, P. R. M., Mainzer, A. K.,  et al., 2010, AJ, 140, 1868 \\
Yor2000 & York, D. G., Adelman, J., Anderson, J. E., Jr., Anderson, S. F.,  et al., 2000, AJ, 120, 1579\\
Yur2011 & Yurchenko, S. N., Barber, R. J. \& Tennyson, J., 2011, MNRAS, 413, 1828\\
Yur2014 & Yurchenko, S.N. \& Tennyson, J., 2014, MNRAS, 440, 1649\\
\end{tabular}
}
\end{table}

\end{document}